\documentclass[10pt,twocolumn]{article}
\usepackage[a4paper,top=2cm,bottom=2.5cm,left=1.5cm,right=1.5cm]{geometry}
\usepackage[utf8]{inputenc}
\usepackage{amsmath,amsthm,amssymb}
\usepackage{graphicx}
\usepackage{array}
\usepackage{url}
\usepackage{bm}
\usepackage{color}
\usepackage{braket}
\usepackage{algorithm,algpseudocode,float}
\usepackage{lipsum}
\usepackage{caption}

\makeatletter
\newenvironment{breakablealgorithm}
  {
   \begin{center}
     \refstepcounter{algorithm}
     \hrule height.8pt depth0pt \kern6pt
     \renewcommand{\caption}[2][\relax]{
       {\raggedright\textbf{\ALG@name~\thealgorithm} ##2\par}%
       \ifx\relax##1\relax 
         \addcontentsline{loa}{algorithm*}{\protect\numberline{\thealgorithm}##2}%
       \else 
         \addcontentsline{loa}{algorithm*}{\protect\numberline{\thealgorithm}##1}%
       \fi
       \kern4pt\hrule\kern4pt
     }
  }{
     \kern8pt\hrule\relax
   \end{center}
  }

\newcommand{\titlepagegeometry}{\newgeometry{top=3cm,bottom=2.5cm,left=1.5cm,right=1.5cm}}

\title{A generalized variational quantum linear solver on photonic platform}

\author{
Kang Gao\textsuperscript{1,$\ast$,$\dagger$},
Zhao-An Wang\textsuperscript{1,$\dagger$},
Ze-Guo Wang\textsuperscript{1},
Mao-Mao Huang\textsuperscript{1},
Hai Wei\textsuperscript{1,2,$\ast$},
Kai Wen\textsuperscript{1,2,$\ast$} \\
\textsuperscript{1}\textit{Quantum Science Center of Guangdong-Hong Kong-Macao Greater Bay Area, Shenzhen 518045, China} \\
\textsuperscript{2}\textit{Beijing QBoson Quantum Technology Co., Ltd., Beijing 100015, China} 
}

\begin{document}
\captionsetup[figure]{labelfont={bf},labelformat={default},labelsep=period,name={Fig.}}

\titlepagegeometry
\maketitle
\restoregeometry

\renewcommand{\thefootnote}{\fnsymbol{footnote}}
\footnotetext[1]{\textit{Email: gaokang@quantumsc.cn/weihai@quantumsc.cn/\\wenkai@quantumsc.cn}}
\footnotetext[2]{\textit{These authors contributed equally to this work.}}
\renewcommand{\thefootnote}{\arabic{footnote}}

\begin{abstract}
Based on a photonic computing platform, we experimentally validate a generalized variational quantum linear solver (VQLS) by systematically solving four-dimensional linear equation systems across different fields. In the complex field, in addition to solving non-singular systems that admit a unique solution, we investigate ill-conditioned problems arising from singularity—--an issue frequently encountered in practical applications. To tackle these challenges, we introduce perturbation terms, a treatment inspired by Tikhonov regularization, and develop an algorithm capable of handling a wide range of systems. Furthermore, we extend the VQLS to the finite field $\mathbb{F}_2$ by redesigning the cost function to incorporate modulo 2 and imposing several constraints on the solution vector. This modulo 2 VQLS is inherently free from singularity. It is adapt to stabilizer coding theory and may find applications in areas such as decoders and the design of quantum gate sequences. Therefore, our work demonstrates the practical potential of VQLS in quantum computing, providing a solid experimental foundation and methodological guidance for its real-world applications.
\end{abstract}


\section{Introduction}

Quantum computing has attracted growing attention as a new computing method \cite{graziano2010,biamonte2017}. In recent years, optimization-referring here to finding the best solution from a set of possible ones-has become one of the most significant challenges for quantum computing \cite{wei2016,Li2019,gao2021,rebentrost2019}. Quantum computers are expected to address this by achieving exponential acceleration, meaning much faster problem-solving compared to classical computers \cite{li2021optimizing,Broughton2020,BravoPrieto2023,Du2019,Wang2018}. The variational quantum algorithm, which uses adjustable parameters in quantum circuits to minimize or maximize a cost function, is a practical quantum computing strategy \cite{peruzzo2014,cerezo2021,farhi2014quantum}. It effectively uses quantum resources and promises advances in optimization problems, such as the MaxCut problem and the molecular heat conduction model \cite{amaro2022,wei2023}. Photonic computing systems offer a strong platform for implementing variational quantum algorithms. Their effectiveness has been validated in multiple experiments \cite{agresti2025demonstration,cimini2024variational,lee2024photonic,stornati2024variational,nielsen2025variational,enomoto2023continuous}, one well-known practical example of which is the calculation of the ground-state molecular energy in quantum chemistry. \cite{peruzzo2014,kim2024qudit,baldazzi2025four,kandala2017hardware}. 

Variational quantum linear solver (VQLS), proposed by C. Bravo-Prieto in 2019 \cite{bravo2023variational}, has emerged as a compelling and methodologically productive topic in quantum computing research \cite{trahan2023variational,huang2021near,rao2024performance,ali2023performance,ghisoni2024shadow}. Numerous simulation-based studies \cite{luo2024variational,turati2024empirical,balducci2024solving} have explored various ansatz structures and optimization strategies—--such as Hamiltonian morphing \cite{xu2021variational}—--to improve circuit design and reduce resource costs. However, experimental tests on real quantum hardware remain rare. This gap is particularly pronounced in photonic quantum computing systems. Consequently, the performance of the solution process and the numerical accuracy of the resulting linear equation solutions have not yet been systematically investigated.

In this work, we demonstrate a generalized variational quantum linear solver on photonic platform. Firstly, we discuss solving linear systems in the complex field and consider the circumstances where the system has a unique solution, infinitely many solutions as well as no solution. For the latter two, the coefficient matrix is singular which may lead to incorrect solution in VQLS. Enlightened by Tikhonov regularization \cite{tikhonov1963} in classical machine learning, we introduce a perturbation term to address singularity, which helps check for solution existence and find approximations if necessary. These discussions are combined into an algorithm that is capable of handling a majority of linear systems. Additionally, we extend the scope of VQLS to a finite field $\mathbb{F}_2$ \cite{aboumrad2024}. By designing a new cost function that reflects modulo 2 via a sinusoidal function and includes several restrictions, we manage to obtain a correct solution in this finite field, which is possibly applicable in the membership problem that one may encounter in the field of stabilizer code.
The present work seeks to bridge the gap between theoretical analysis and real-device realization, towards the first step of the practical application of VQLS in the Noisy Intermediate-Scale Quantum computing (NISQ) era.

\begin{table*}[ht]
\begin{breakablealgorithm}
    \caption{The algorithm for solving a linear system with a variational quantum linear solver (VQLS) in the complex field}\label{alg:VQLS}
    \begin{algorithmic}
        \Require
            The coefficient matrix $\bm{A}$, the unnormalized constant vector $\ket{\tilde{\bm{b}}}$ of the linear system and thresholds $\zeta_{1,2}$
        \Ensure
            The unique or one of the (approximated) solution of the system ($\ket{\tilde{\bm{x}}}=\Gamma\ket{\bm{x}}$), or determine that there is no solution
        \State Calculate the Hamiltonian $\bm{H}$ and the decomposition coefficients $\bm{c}_k$ in Eq.\,\ref{eq:HG&cost}
        \State Run the optimization process and find a vector $\ket{\bm{x}}$ that minimizes the cost value, and calculate $\Gamma$
        \State Calculate the deviation $\mathcal{D}=||\bm{A}\ket{\tilde{\bm{x}}}-\ket{\tilde{\bm{b}}}||^2$
        \If {$\mathcal{D} \leq \zeta_1$}
            \State \Return {$\ket{\tilde{\bm{x}}}$ is the unique or one of the infinite number of solutions of the system}
        \Else 
        \State Introduce a positive real value $\delta$ and modify the input system as well as the cost function
        \State Run the optimization process and find a vector $\ket{\bm{x}}$ that minimizes the cost value, and calculate $\Gamma$
        \State Calculate the deviation $\mathcal{D}=||\bm{A}\ket{\tilde{\bm{x}}}-\ket{\tilde{\bm{b}}}||^2$
            \If {$\mathcal{D} \leq \zeta_2$}
                \State \Return {There are infinitely many solutions and $\ket{\tilde{\bm{x}}}$ is one of the approximated solutions}
            \Else
                \State \Return{There is no solution}
            \EndIf
        \EndIf
    \end{algorithmic}
\end{breakablealgorithm}
\end{table*}

\section{Theoretical framework}
The principle of variational quantum algorithms can be summarized as follows: initialize a set of parameters, prepare the corresponding quantum state; evaluate a carefully designed cost function which is typically expressed as the expectation value of a Hamiltonian that quantifies solution accuracy; optimize the classical parameters to minimize this cost function. This underlying philosophy is naturally inherited by the VQLS, which we discuss successively in the complex field and over the finite field $\mathbb{F}_2$ in succession. The system of linear equations is expressed as
\begin{equation}
\bm{A}|\bm{\tilde{x}}\rangle = |\bm{\tilde{b}}\rangle,
\label{eq:linear}
\end{equation}
where $\bm{A}$, $\ket{\bm{\tilde{b}}}$ and $\ket{\bm{\tilde{x}}}$ represent the known coefficient matrix, the unnormalized constant vector, and the unknown unnormalized target vector, respectively.

\subsection{VQLS in the complex field}
\noindent\textbf{Non-singular $\bm{A}$}

In the standard setting, we assume that the coefficient matrix $\bm{A}$ is of full rank (i.e., non‑singular), so that the linear system admits a unique solution. The Hamiltonian and the cost function are defined as follows:
\begin{equation}
\begin{aligned}
\bm{H} &= \bm{A}^\dagger(\braket{\bm{\tilde{b}}|\bm{\tilde{b}}}\mathbf{I} - |\bm{b}\rangle\langle\bm{b}|)\bm{A} =  \sum_k{\bm{c}}_{k}\bm{U}_{k},\\
Cost(\bm{\theta}) &= f\left(\langle \bm{x}|\bm{H} |\bm{x}\rangle\right),
\end{aligned}
\label{eq:HG&cost}
\end{equation}
where $\mathbf{I}$ is the identity matrix, $\ket{\bm{b}}$ and $\ket{\bm{x}}$ have the unity norm, $\ket{\tilde{\bm{x}}}=\Gamma\ket{\bm{x}}$ and $\Gamma$ is a global constant. The $\bm{\theta}$ (which correspond to waveplate angles in our photonic implementation) are to be optimized. The Hamiltonian is expressed as a superposition of orthogonal Hermitian operators: the matrices ${\bm{U}}_{k}$ form a basis consisting of tensor products of Pauli matrices, and the coefficients are given by ${\bm{c}}_{k}=\mathrm{Tr}[\bm{H} \bm{U}_k]$. 

$f(\cdot)$ is a nonlinear activation function that transforms the input to accelerate convergence. The expected value of the Hamiltonian is experimentally measurable and the cost function reaches its minimum, zero, exactly when the unique solution is obtained.

\noindent\textbf{Singular $\bm{A}$}

We now turn to the case where $\bm{A}$ is singular, so that the linear system may have either infinitely many solutions or have no solution, a situation that is problematic for both classical machine learning and VQLS. Consider, for instance, a system with infinitely many solutions, whose solution vectors take the form $\ket{\tilde{\bm{x}}}=(x_1,0,0,1-x_1)^\mathrm{T}$. Unlike the non‑singular case, here the cost can vanish while the final $\ket{\bm{x}}$ is not necessarily a true solution. As a matter of fact, singularity of $\bm{A}$ renders $\bm{H}$ ill-conditioned: for any complex numbers $z_1$ and $z_2$, the vector $(z_1,0,0,z_2)^\mathrm{T}$ yields a zero cost, yet most such vectors are not solutions. Hence, a direct application of the previous method is prone to incorrect results. Systems without any solution suffer from analogous difficulties.

To tackle this issue, we draw inspiration from the treatment of singular $\bm{A}$ in classical machine learning and incorporate Tikhonov regularization into VQLS \cite{tikhonov1963}. Minimizing the deviation $\mathcal{D}=||\bm{A}\ket{\tilde{\bm{x}}}-\ket{\tilde{\bm{b}}}||^2=(\bm{A}\ket{\tilde{\bm{x}}}-\ket{\tilde{\bm{b}}})^\dagger(\bm{A}\ket{\tilde{\bm{x}}}-\ket{\tilde{\bm{b}}})$ is equivalent to solving the normal equation $\bm{A}^\dagger\bm{A}\ket{\tilde{\bm{x}}}=\bm{A}^\dagger\ket{\tilde{\bm{b}}}$, which follows from setting $\partial\mathcal{D}/\partial\ket{\tilde{\bm{x}}}=0$. In the context of classical machine learning, a standard regularization approach adds a perturbation term $\delta \mathbf{I}$ (with $\delta>0$ for simplicity), rendering it non‑singular and yielding an approximate solution. For VQLS, we can adopt a similar strategy: replace $\bm{A}$ with $\bm{A}^\prime = \bm{A}^\dagger \bm{A}+\delta\mathbf{I}$ and $\ket{\bm{b}}$ with $\ket{\bm{b^\prime}} = \bm{A}^\dagger\ket{\bm{b}}$ in Eq.\,\ref{eq:linear}, and then apply the cost function defined in Eq.\,\ref{eq:HG&cost}.
A larger $\delta$ implies $\bm{A}$ makes the modified matrix further from singularity and enhances robustness against noise, but at the expense of potentially lower accuracy of the approximated solution. In practice, we may accept the approximate solution provided that the deviation $\mathcal{D}$ does not exceed a preset threshold $\zeta$.

We have thus covered the main scenarios for VQLS: unique solution, infinitely many solutions, and no solution. The complete algorithmic procedure is summarized in pseudocode in Algorithm.\,\ref{alg:VQLS}.

\begin{figure*}[ht]
    \centering
    \includegraphics[width=\linewidth]{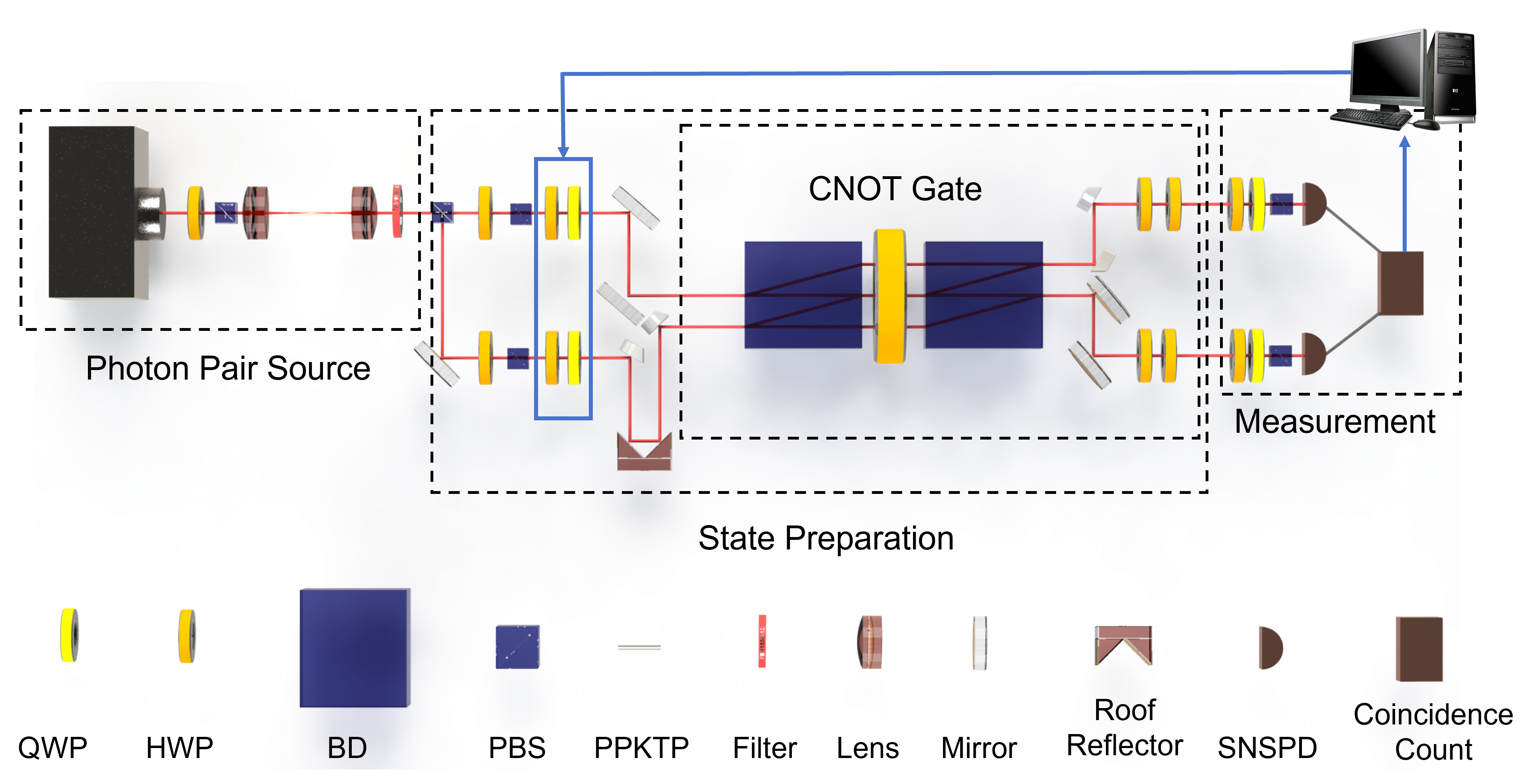}
    \caption{\textbf{Experimental setup.} Telecom-band photon pairs are generated and pass through a state preparation process including a CNOT gate and are subjected to coincidence measurement. The measurement results are analyzed and exploited to update the angles of waveplates enclosed in the blue frame. QWP: quarter waveplate, HWP: half waveplate, BD: beam displacer, PBS: polarization beam splitter, SNSPD: superconducting nanowire single photon detector.}
    \label{fig:setup}
\end{figure*}

\subsection{Modulo 2 VQLS}

In this section, we consider the VQLS in the finite field $\mathbb{F}_2$, where all entries of $\bm{A}$, $\ket{\tilde{\bm{x}}}$ and $\ket{\tilde{\bm{b}}}$ are restricted to $\{0,1\}$ and arithmetic follows modulo 2.  In this setting, rather than constructing a single Hamiltonian from the entire $\bm{A}$ and $\ket{\tilde{\bm{b}}}$ as in Eq.\,\ref{eq:HG&cost}, we treat each linear equation $\braket{\bm{a_k}|\tilde{\bm{x}}}=\tilde{b}_k$ individually. For each row $k$, we define the Hamiltonian $\bm{H_k}=\ket{\bm{a_k}}\bra{\bm{a_k}}$, where $\bra{\bm{a_k}}$ is the $k$-th row of $\bm{A}$ and $\tilde{b}_k$ is the corresponding entry of $\ket{\tilde{\bm{b}}}$. Notably, singularity of $\bm{A}$ no longer poses a difficulty: when no solution exists, the cost value will stop decreasing above zero. In the complex-field cases discussed earlier, vanishing cost does not guarantee a correct solution because the phases of the entries of $\ket{\bm{x}}$ are indeterminate. In modulo 2 VQLS, however, this issue is circumvented, and with suitable constraints we can reliably obtain the correct solution. The cost function is defined as:
\begin{equation}
\begin{aligned}
J_k &= \sin^2 \left[\frac{\pi}{2} \left(\gamma\sqrt{\braket{\bm{x}|\bm{H}_k|\bm{x}}}+\tilde{b}_k\right)\right],\\
Cost(\bm{\theta},\gamma) &= f(J_1, J_2,...) + \lambda_{1} \sum_{i,j} |\tilde{x}_i \tilde{x}_j|^2 (|\tilde{x}_i|^2-|\tilde{x}_j|^2)^2\\
     &+ \lambda_2 g(\gamma) + \lambda_3 \sin^2(\pi \gamma).
\end{aligned}
\label{eq:costmod2}
\end{equation}

The design of this cost function incorporates several key ideas. First, to encode modulo 2 arithmetic within the real numbers, the main term is chosen as a sinusoidal function with period 2.
 
Second, we require $\ket{{\bm{x}}}$ to be a real vector, a condition readily achievable in photonic platforms. Moreover, each entry of $\ket{\tilde{\bm{x}}}$ must be either zero or have the same modulus as all other nonzero entries; note that $1$ and $-1$ are equivalent under modulo 2. This constraint can be enforced by simply measuring in the $Z$ basis, and its strength is controlled by the real parameter $\lambda_1$.
Third, while $\ket{\bm{x}}$ is normalized to unity, the actual solution vector $\ket{\tilde{\bm{x}}}$ has a norm equal to the square root of the number of nonzero entries. Therefore, we introduce the global constant $\gamma=|\Gamma|^2$ as an additional optimization variable alongside the waveplate angles. Two restrictions are imposed on $\gamma$: $g(\gamma)$ is a ``pit‑shaped'' function that is nearly zero when $\gamma \in [1,\mathrm{dim}(\ket{\bm{x}})]$ but rises sharply outside this interval; $\gamma$ must be an integer. The range constraint and the integer constraint are weighted by $\lambda_2$ and $\lambda_3$, respectively, with $\lambda_3$ being activated only after many iterations when the descent becomes considerably slow.

\section{Experimental setup and results}
\noindent\textbf{Setup}

The experimental setup is depicted in Fig.\,\ref{fig:setup}. A pair of telecom-band photons are generated from a type-II periodically poled Potassium Titanyl Phosphate (PPKTP) crystal, and the qubits are encoded in the polarization. The target vector $\ket{\bm{x}}$ is then prepared via several half waveplates (HWPs) and quarter waveplates (QWPs), together with a CNOT gate composed of beam displacers (BDs) and HWPs \cite{OBrien2003}. The angles of a pair of HWP-QWP combination (highlighted in yellow and orange in Fig.\,\ref{fig:setup}), denoted by the entries of $\bm{\theta}$, serve as the variational parameters of the cost function and are updated during the optimization. Finally, the photons are collected and subjected to coincidence measurement, projecting $\ket{\bm{x}}$ onto the bases of $\bm{U}_k$ and yielding the expected value $\braket{\bm{x}|\bm{U}_k|\bm{x}}$.

In each optimization cycle, we first obtain a set of trial cost values by slightly adjusting each component of $\bm{\theta}$ individually. We then compute the gradients in each direction and update $\bm{\theta}$ using the Adam gradient descent algorithm.

\begin{figure*}[!ht]
    \centering
    \includegraphics[width=1\linewidth]{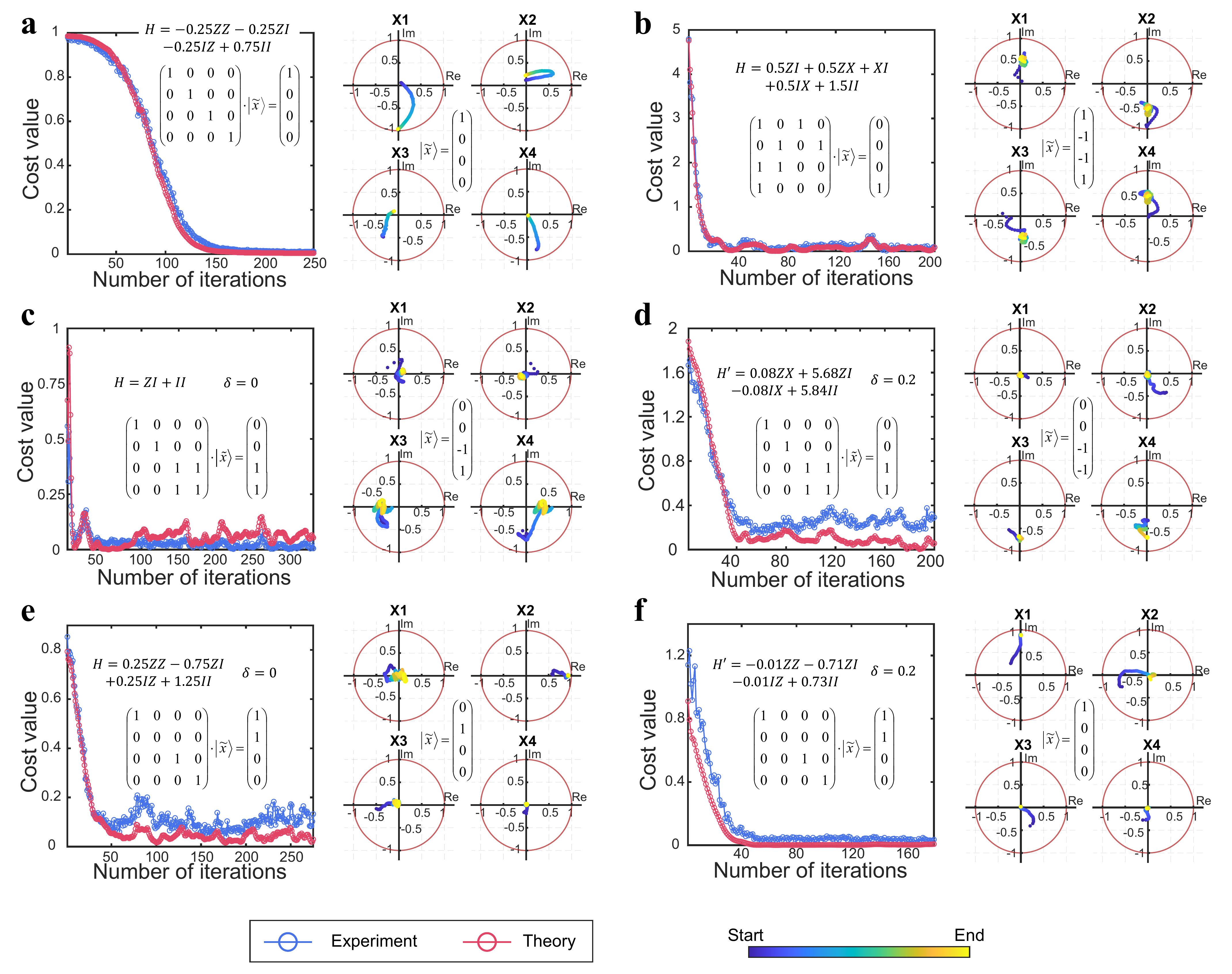}
    \caption{\textbf{Demonstration of VQLS in complex field.} \textbf{a-b}: The coefficient matrix $\bm{A}$ is non-singular and the linear system has a unique solution. \textbf{c-d}: $\bm{A}$ is singular and the system has infinitely many solutions. The system is solved without (\textbf{c}) and with Tikhonov regularization ($\delta=0.2$, \textbf{d}), respectively. An approximated solution can be obtained in the latter circumstances. \textbf{e-f}. The system has no solution and is solved without (\textbf{e}) and with Tikhonov regularization ($\delta=0.2$, \textbf{f}), respectively. The final $\ket{\tilde{\bm{x}}}$ in neither circumstances is verified the solution of the system. The left subplot in each panel is the optimization process of the cost value. The blue and red points and lines represent experimental cost value and the theoretical results from the same set of $\bm{\theta}$. The right subplot exhibits the trajectories of the entries of $\ket{\bm{x}}$ corresponding to $\bm{\theta}$ and the color bar indicates the direction of the trajectories.}
    \label{fig:complex}
\end{figure*}

\noindent\textbf{Results}

Fig.\,\ref{fig:complex} presents the results of applying VQLS in the complex field. In Fig.\,\ref{fig:complex}a-b, correspond to a non‑singular $\bm{A}$. We show the evolution of the cost value and the trajectories of the trial $\ket{\bm{x}}$ entries in the complex plane, as determined from the waveplate angles. In the cost‑value plots, the blue and red lines represent the experimental results and the corresponding theoretical predictions, respectively. The cost value steadily decreases toward zero, and the trial $\ket{\bm{x}}$ converges to the correct solution. These results confirm that our experimental setup can successfully handle the basic non‑singular case.

In Fig.\,\ref{fig:complex}c, we show an example where the cost value is optimized to near zero, yet the final $\ket{\tilde{\bm{x}}}$ is not a valid solution. In panel d, we apply Tikhonov regularization with $\delta=0.2$ to the same system as in panel c and present the corresponding results. In this example, the approximated solution of the regularized system shown here remains identical (up to a global constant that does not affect $\ket{\bm{x}}$) to one of the solutions of the original system. These results demonstrate that VQLS successfully finds the correct solution with the aid of $\delta$. Fig.\,\ref{fig:complex}e-f show results for a system with no solution. In both the regularized and original cases, a state $\ket{\bm{x}}$ that minimizes the cost to near zero is obtained. However, upon direct verification, neither of these states is found to be a solution of the original system.

\begin{figure}[!ht]
    \centering
    \includegraphics[width=\linewidth]{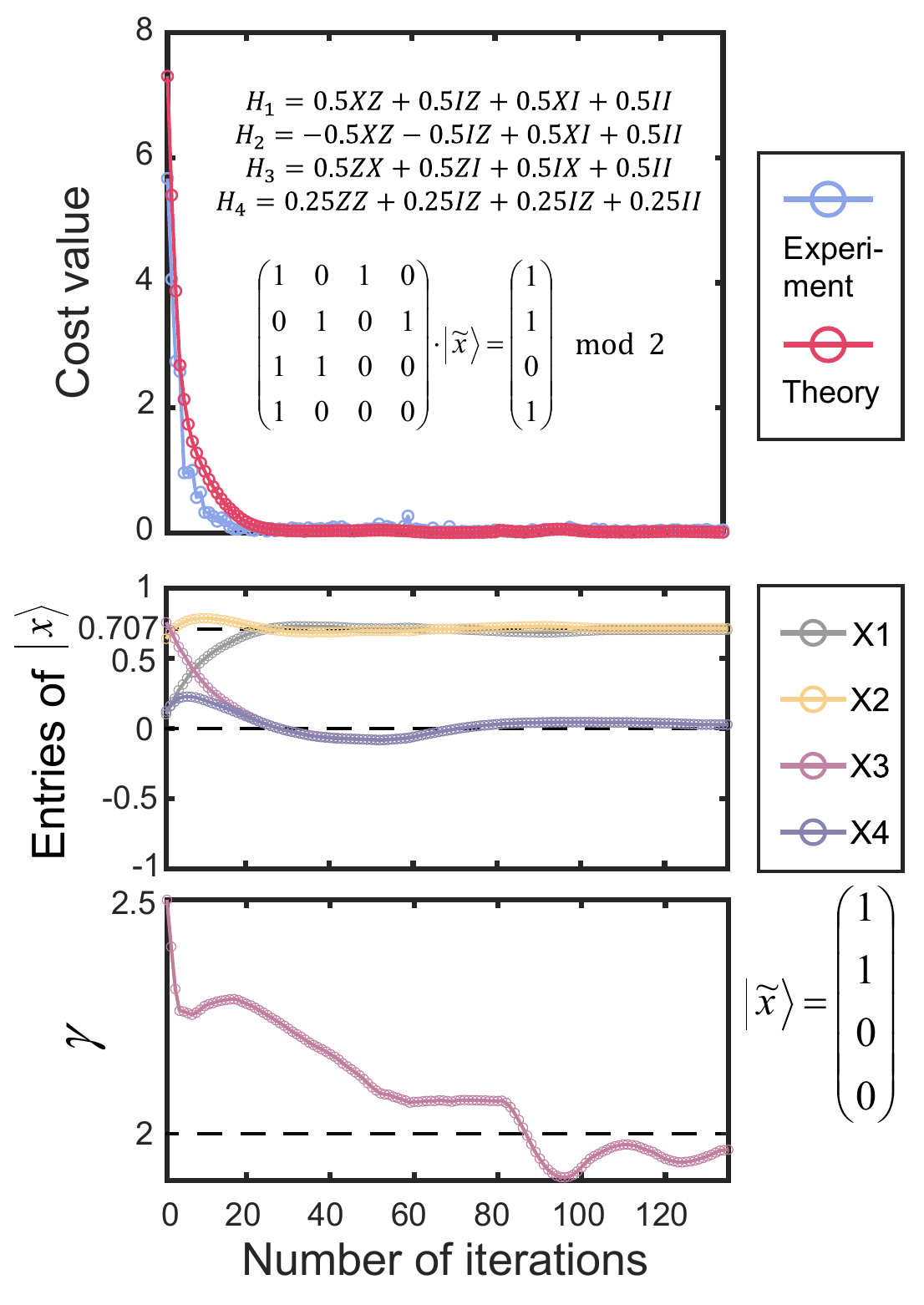}
    \caption{\textbf{Demonstration of Modulo 2 VQLS.} Four Hamiltonians are introduced according to the rows of $\bm{A}$ and restrictions are imposed on the entries of $\ket{\tilde{\bm{x}}}$ via the Pauli measurements. The upper subplot is the optimization process of the cost value. The blue and red points and lines represent experimental cost value and the theoretical results from the same set of $\bm{\theta}$. The lower two subplots exhibit the trajectories of the entries of $\ket{\bm{x}}$ and $\gamma$. Dotted lines indicate the true solution values for both the entries of $\ket{\bm{x}}$ and $\gamma$.}
    \label{fig:mod2}
\end{figure}

Finally, Fig.\,\ref{fig:mod2} presents the results of the modulo 2 VQLS. Since $\ket{\tilde{\bm{x}}}$ is real in the modulo 2 setting, the QWPs can be removed or their angles can be locked to those of the HWPs to eliminate imaginary components. The trajectories are therefore confined to the real axis, and the correct solution is successfully obtained.

\section{Conclusion and further applications}
In conclusion, we have experimentally implemented a four‑dimensional VQLS and demonstrated its validity in both the complex field and the finite field $\mathbb{F}_2$ (Modulo 2 VQLS). In the former situation, we have discussed the singularity of $\bm{A}$ and have proposed a possible approach for handling singular $\bm{A}$ in analogy of classical machine learning. On top of this, we have formed a complete algorithm flow. For the modulo 2 case, we have designed a new cost function that decomposes $\bm{A}$ row by row and imposes several parameter constraints; this approach is not limited by the singularity of $\bm{A}$. By broadening the scope of VQLS, we believe our work may find further applications in a variety of contexts and scientific problems.
 
Potential applications of VQLS in the continuous (including complex) domain include quantum support vector machines, time‑series forecasting, and partial differential equation constrained optimization \cite{yi2023,dimitrijevs2024,surana2025}. These applications may suffer from singularities that lead to issues such as convergence failure, and our method offers a viable way to mitigate these difficulties. Moreover, modulo 2 VQLS is broadly applicable across diverse fields—for instance, in determining the number of solutions to systems of polynomial equations over finite fields for point‑counting in algebraic geometry \cite{harrow2009}, in detecting and characterizing edge states of two‑dimensional topological phases in condensed matter physics \cite{larose2019}, and in solving the membership problem for stabilizer codes within quantum computation. We briefly illustrate this last example below.
 
Consider a group $S=\langle\sigma_{11}\sigma_{12}...,\,\sigma_{21}\sigma_{22}...,\,...\rangle$ consisting of mutually commuting and independent stabilizers, where each $\sigma_{ij}$ a Pauli matrix. The stabilizer code is defined to be the vector space stabilized by $S$, and operators that commute with all the stabilizers but are not members of $S$ are logical operators or logical errors. In the language of parity check, the Pauli matrices $\mathbf{I}$, $X$, $Y$ and $Z$ are mapped to $(0,0)^\mathrm{T}$, $(1,0)^\mathrm{T}$, $(1,1)^\mathrm{T}$, $(0,1)^\mathrm{T}$, respectively, e.g., group element $XYZ\mathbf{I}$ corresponds to $(0,1,1,0,1,1,0,0)^\mathrm{T}$. Multiplication of Pauli matrices is consequently mapped to modulo 2 addition ignoring the extra $\pm i$. Therefore, given $S$ and an operator $M$ commuting to all stabilizers, modulo 2 VQLS allows one to determine whether $M$ is a member of $S$ and how it can be factorized by the stabilizers or logical operators, consequently assess its effect on the logical state. As an example, results in Fig.\,\ref{fig:mod2} can be interpreted with $S=\langle YZ, ZX \rangle$ and $M=XY$. We note that in the modulo‑2 VQLS setting, $\bm{A}$ need not be square, and the last two columns may be discarded.

\section*{Acknowledgments}
This work is supported by the Research Initiation Fund of the Quantum Science Center of the Guangdong–Hong Kong–Macao Greater Bay Area (Grant No. QD2305001) and the Station of Quantum Materials.
Z.-G. W. further acknowledges support from the National Natural Science Foundation of China (Grant No. 62501522) and the Guangdong Provincial Quantum Science Strategic Initiative (Grant No. GDZX2503005).

\bibliographystyle{unsrt}
\bibliography{reference}

@article{graziano2010,
  title={LMI techniques for optimization over polynomials in control: a survey},
  author={Chesi, Graziano},
  journal={IEEE transactions on Automatic Control},
  volume={55},
  number={11},
  pages={2500--2510},
  year={2010},
  publisher={IEEE}
}

@article{biamonte2017,
  title={Quantum machine learning},
  author={Biamonte, Jacob and Wittek, Peter and Pancotti, Nicola and Rebentrost, Patrick and Wiebe, Nathan and Lloyd, Seth},
  journal={Nature},
  volume={549},
  number={7671},
  pages={195--202},
  year={2017},
  publisher={Nature Publishing Group UK London}
}

@article{wei2016,
  title={Duality quantum computer and the efficient quantum simulations},
  author={Wei, Shi-Jie and Long, Gui-Lu},
  journal={Quantum Information Processing},
  volume={15},
  number={3},
  pages={1189--1212},
  year={2016},
  publisher={Springer}
}

@article{li2019,
  title={Measuring holographic entanglement entropy on a quantum simulator},
  author={Li, Keren and Han, Muxin and Qu, Dongxue and Huang, Zichang and Long, Guilu and Wan, Yidun and Lu, Dawei and Zeng, Bei and Laflamme, Raymond},
  journal={npj Quantum Information},
  volume={5},
  number={1},
  pages={30},
  year={2019},
  publisher={Nature Publishing Group UK London}
}

@article{gao2021,
  title={Quantum gradient algorithm for general polynomials},
  author={Gao, Pan and Li, Keren and Wei, Shijie and Gao, Jiancun and Long, Guilu},
  journal={Physical Review A},
  volume={103},
  number={4},
  pages={042403},
  year={2021},
  publisher={APS}
}

@article{rebentrost2019,
  title={Quantum gradient descent and Newton’s method for constrained polynomial optimization},
  author={Rebentrost, Patrick and Schuld, Maria and Wossnig, Leonard and Petruccione, Francesco and Lloyd, Seth},
  journal={New Journal of Physics},
  volume={21},
  number={7},
  pages={073023},
  year={2019},
  publisher={IOP Publishing}
}

@article{li2021optimizing,
  title={Optimizing a polynomial function on a quantum processor},
  author={Li, Keren and Wei, Shijie and Gao, Pan and Zhang, Feihao and Zhou, Zengrong and Xin, Tao and Wang, Xiaoting and Rebentrost, Patrick and Long, Guilu},
  journal={npj Quantum Information},
  volume={7},
  number={1},
  pages={16},
  year={2021},
  publisher={Nature Publishing Group UK London}
}

@article{broughton2020,
  title={Tensorflow quantum: A software framework for quantum machine learning},
  author={Broughton, Michael and Verdon, Guillaume and McCourt, Trevor and Martinez, Antonio J and Yoo, Jae Hyeon and Isakov, Sergei V and Massey, Philip and Halavati, Ramin and Niu, Murphy Yuezhen and Zlokapa, Alexander and others},
  journal={arXiv preprint arXiv:2003.02989},
  year={2020}
}

@article{BravoPrieto2023,
  title={Variational quantum linear solver},
  author={Bravo-Prieto, Carlos and LaRose, Ryan and Cerezo, Marco and Subasi, Yigit and Cincio, Lukasz and Coles, Patrick J},
  journal={Quantum},
  volume={7},
  pages={1188},
  year={2023},
  publisher={Verein zur F{\"o}rderung des Open Access Publizierens in den Quantenwissenschaften}
}

@inproceedings{du2019,
  title={Gradient descent finds global minima of deep neural networks},
  author={Du, Simon and Lee, Jason and Li, Haochuan and Wang, Liwei and Zhai, Xiyu},
  booktitle={International conference on machine learning},
  pages={1675--1685},
  year={2019},
  organization={PMLR}
}

@article{wang2018,
  title={Insensitive stochastic gradient twin support vector machines for large scale problems},
  author={Wang, Zhen and Shao, Yuan-Hai and Bai, Lan and Li, Chun-Na and Liu, Li-Ming and Deng, Nai-Yang},
  journal={Information sciences},
  volume={462},
  pages={114--131},
  year={2018},
  publisher={Elsevier}
}

@article{peruzzo2014,
  title={A variational eigenvalue solver on a photonic quantum processor},
  author={Peruzzo, Alberto and McClean, Jarrod and Shadbolt, Peter and Yung, Man-Hong and Zhou, Xiao-Qi and Love, Peter J and Aspuru-Guzik, Al{\'a}n and O’brien, Jeremy L},
  journal={Nature communications},
  volume={5},
  number={1},
  pages={4213},
  year={2014},
  publisher={Nature Publishing Group UK London}
}

@article{cerezo2021,
  title={Variational quantum algorithms},
  author={Cerezo, Marco and Arrasmith, Andrew and Babbush, Ryan and Benjamin, Simon C and Endo, Suguru and Fujii, Keisuke and McClean, Jarrod R and Mitarai, Kosuke and Yuan, Xiao and Cincio, Lukasz and others},
  journal={Nature Reviews Physics},
  volume={3},
  number={9},
  pages={625--644},
  year={2021},
  publisher={Nature Publishing Group UK London}
}

@article{OBrien2003,
  title={Demonstration of an all-optical quantum controlled-NOT gate},
  author={O'Brien, Jeremy L and Pryde, Geoffrey J and White, Andrew G and Ralph, Timothy C and Branning, David},
  journal={Nature},
  volume={426},
  number={6964},
  pages={264--267},
  year={2003},
  publisher={Nature Publishing Group UK London}
}

@article{amaro2022,
  title={Filtering variational quantum algorithms for combinatorial optimization},
  author={Amaro, David and Modica, Carlo and Rosenkranz, Matthias and Fiorentini, Mattia and Benedetti, Marcello and Lubasch, Michael},
  journal={Quantum Science \& Technology},
  volume={7},
  number={1},
  pages={015021},
  year={2022},
  publisher={IOP Publishing}
}

@article{harrow2009,
  title={Quantum algorithm for linear systems of equations},
  author={Harrow, Aram W and Hassidim, Avinatan and Lloyd, Seth},
  journal={Physical review letters},
  volume={103},
  number={15},
  pages={150502},
  year={2009},
  publisher={APS}
}

@article{larose2019,
  title={Variational quantum state diagonalization},
  author={LaRose, Ryan and Tikku, Arkin and O’Neel-Judy, {\'E}tude and Cincio, Lukasz and Coles, Patrick J},
  journal={npj Quantum Information},
  volume={5},
  number={1},
  pages={57},
  year={2019},
  publisher={Nature Publishing Group UK London}
}

@article{aboumrad2024,
  title={Mod2VQLS: A variational quantum algorithm for solving systems of linear equations modulo 2},
  author={Aboumrad, Willie and Widdows, Dominic},
  journal={Applied Sciences},
  volume={14},
  number={2},
  pages={792},
  year={2024},
  publisher={MDPI}
}

@article{wei2023,
  title={A quantum algorithm for heat conduction with symmetrization},
  author={Wei, Shi-Jie and Wei, Chao and Lv, Peng and Shao, Changpeng and Gao, Pan and Zhou, Zengrong and Li, Keren and Xin, Tao and Long, Gui-Lu},
  journal={Science Bulletin},
  volume={68},
  number={5},
  pages={494--502},
  year={2023},
  publisher={Elsevier}
}

@article{farhi2014quantum,
  title={A quantum approximate optimization algorithm},
  author={Farhi, Edward and Goldstone, Jeffrey and Gutmann, Sam},
  journal={arXiv preprint arXiv:1411.4028},
  year={2014}
}

@article{kim2024qudit,
  title={Qudit-based variational quantum eigensolver using photonic orbital angular momentum states},
  author={Kim, Byungjoo and Hu, Kang-Min and Sohn, Myung-Hyun and Kim, Yosep and Kim, Yong-Su and Lee, Seung-Woo and Lim, Hyang-Tag},
  journal={Science Advances},
  volume={10},
  number={43},
  pages={eado3472},
  year={2024},
  publisher={American Association for the Advancement of Science}
}

@article{baldazzi2025four,
  title={Four-qubit variational algorithms in silicon photonics with integrated entangled photon sources},
  author={Baldazzi, Alessio and Sanna, Matteo and Borghi, Massimo and Azzini, Stefano and Pavesi, Lorenzo},
  journal={npj Quantum Information},
  volume={11},
  number={1},
  pages={107},
  year={2025},
  publisher={Nature Publishing Group UK London}
}

@article{agresti2025demonstration,
  title={Demonstration of hardware efficient photonic variational quantum algorithm},
  author={Agresti, Iris and Paul, Koushik and Schiansky, Peter and Steiner, Simon and Yin, Zhenghao and Pentangelo, Ciro and Piacentini, Simone and Crespi, Andrea and Ban, Yue and Ceccarelli, Francesco and others},
  journal={Physical Review Research},
  volume={7},
  number={4},
  pages={043021},
  year={2025},
  publisher={APS}
}

@article{cimini2024variational,
  title={Variational quantum algorithm for experimental photonic multiparameter estimation},
  author={Cimini, Valeria and Valeri, Mauro and Piacentini, Simone and Ceccarelli, Francesco and Corrielli, Giacomo and Osellame, Roberto and Spagnolo, Nicol{\`o} and Sciarrino, Fabio},
  journal={npj Quantum Information},
  volume={10},
  number={1},
  pages={26},
  year={2024},
  publisher={Nature Publishing Group UK London}
}

@article{lee2024photonic,
  title={Photonic variational quantum eigensolver using entanglement measurements},
  author={Lee, Jinil and Song, Wooyeong and Lee, Donghwa and Kim, Yosep and Lee, Seung-Woo and Lim, Hyang-Tag and Jung, Hojoong and Han, Sang-Wook and Kim, Yong-Su},
  journal={Quantum Science and Technology},
  volume={9},
  number={4},
  pages={045028},
  year={2024},
  publisher={IOP Publishing}
}

@article{stornati2024variational,
  title={Variational quantum simulation using non-Gaussian continuous-variable systems},
  author={Stornati, Paolo and Acin, Antonio and Chabaud, Ulysse and Dauphin, Alexandre and Parigi, Valentina and Centrone, Federico},
  journal={Physical Review Research},
  volume={6},
  number={4},
  pages={043212},
  year={2024},
  publisher={APS}
}

@article{nielsen2025variational,
  title={Variational quantum algorithm for enhanced continuous variable optical phase sensing},
  author={Nielsen, Jens AH and Kicinski, Mateusz J and Arge, Tummas N and Vijayadharan, Kannan and Foldager, Jonathan and Borregaard, Johannes and Meyer, Johannes Jakob and Neergaard-Nielsen, Jonas S and Gehring, Tobias and Andersen, Ulrik L},
  journal={npj Quantum Information},
  volume={11},
  number={1},
  pages={70},
  year={2025},
  publisher={Nature Publishing Group UK London}
}

@article{enomoto2023continuous,
  title={Continuous-variable quantum approximate optimization on a programmable photonic quantum processor},
  author={Enomoto, Yutaro and Anai, Keitaro and Udagawa, Kenta and Takeda, Shuntaro},
  journal={Physical Review Research},
  volume={5},
  number={4},
  pages={043005},
  year={2023},
  publisher={APS}
}

@article{kandala2017hardware,
  title={Hardware-efficient variational quantum eigensolver for small molecules and quantum magnets},
  author={Kandala, Abhinav and Mezzacapo, Antonio and Temme, Kristan and Takita, Maika and Brink, Markus and Chow, Jerry M and Gambetta, Jay M},
  journal={nature},
  volume={549},
  number={7671},
  pages={242--246},
  year={2017},
  publisher={Nature Publishing Group}
}

@article{bravo2023variational,
  title={Variational quantum linear solver},
  author={Bravo-Prieto, Carlos and LaRose, Ryan and Cerezo, Marco and Subasi, Yigit and Cincio, Lukasz and Coles, Patrick J},
  journal={Quantum},
  volume={7},
  pages={1188},
  year={2023},
  publisher={Verein zur F{\"o}rderung des Open Access Publizierens in den Quantenwissenschaften}
}

@article{trahan2023variational,
  title={A variational quantum linear solver application to discrete finite-element methods},
  author={Trahan, Corey Jason and Loveland, Mark and Davis, Noah and Ellison, Elizabeth},
  journal={Entropy},
  volume={25},
  number={4},
  pages={580},
  year={2023},
  publisher={MDPI}
}

@article{huang2021near,
  title={Near-term quantum algorithms for linear systems of equations with regression loss functions},
  author={Huang, Hsin-Yuan and Bharti, Kishor and Rebentrost, Patrick},
  journal={New Journal of Physics},
  volume={23},
  number={11},
  pages={113021},
  year={2021},
  publisher={IOP Publishing}
}

@article{rao2024performance,
  title={Performance study of variational quantum linear solver with an improved ansatz for reservoir flow equations},
  author={Rao, Xiang},
  journal={Physics of Fluids},
  volume={36},
  number={4},
  year={2024},
  publisher={AIP Publishing}
}

@article{ali2023performance,
  title={Performance study of variational quantum algorithms for solving the Poisson equation on a quantum computer},
  author={Ali, Mazen and Kabel, Matthias},
  journal={Physical Review Applied},
  volume={20},
  number={1},
  pages={014054},
  year={2023},
  publisher={APS}
}

@article{ghisoni2024shadow,
  title={Shadow quantum linear solver: A resource efficient quantum algorithm for linear systems of equations},
  author={Ghisoni, Francesco and Scala, Francesco and Bajoni, Daniele and Gerace, Dario},
  journal={arXiv preprint arXiv:2409.08929},
  year={2024}
}

@article{luo2024variational,
  title={Variational quantum linear solver-based combination rules in Dempster--Shafer theory},
  author={Luo, Hao and Zhou, Qianli and Li, Zhen and Deng, Yong},
  journal={Information Fusion},
  volume={102},
  pages={102070},
  year={2024},
  publisher={Elsevier}
}

@article{xu2021variational,
  title={Variational algorithms for linear algebra},
  author={Xu, Xiaosi and Sun, Jinzhao and Endo, Suguru and Li, Ying and Benjamin, Simon C and Yuan, Xiao},
  journal={Science Bulletin},
  volume={66},
  number={21},
  pages={2181--2188},
  year={2021},
  publisher={Elsevier}
}

@article{turati2024empirical,
  title={An empirical analysis on the effectiveness of the variational quantum linear solver},
  author={Turati, Gloria and Marruzzo, Alessia and Dacrema, Maurizio Ferrari and Cremonesi, Paolo},
  journal={arXiv preprint arXiv:2409.06339},
  year={2024}
}

@article{balducci2024solving,
  title={Solving 1D Poisson problem with a variational quantum linear solver},
  author={Balducci, Giorgio Tosti and Chen, Boyang and M{\"o}ller, Matthias and De Breuker, Roeland},
  journal={arXiv preprint arXiv:2412.04938},
  year={2024}
}

@article{dimitrijevs2024,
  title={Exploring Hybrid Quantum-Classical Methods for Practical Time-Series Forecasting},
  author={Dimitrijevs, Maksims and K{\=a}lis, M{\=a}rti{\c{n}}{\v{s}} and Repko, I{\c{l}}ja},
  journal={arXiv preprint arXiv:2412.05615},
  year={2024}
}

@article{surana2025,
  title={Variational quantum framework for partial differential equation constrained optimization},
  author={Surana, Amit and Gnanasekaran, Abeynaya},
  journal={ACM Transactions on Quantum Computing},
  volume={7},
  number={1},
  pages={1--36},
  year={2025},
  publisher={ACM New York, NY}
}

@article{yi2023,
  title={Variational quantum linear solver enhanced quantum support vector machine},
  author={Yi, Jianming and Suresh, Kalyani and Moghiseh, Ali and Wehn, Norbert},
  journal={arXiv preprint arXiv:2309.07770},
  year={2023}
}

@article{tikhonov1963,
  title={Solution of incorrectly formulated problems and the regularization method},
  author={Tikhonov, Andrei N},
  journal={Sov Dok},
  volume={4},
  pages={1035--1038},
  year={1963}
}
\end{document}